\documentclass[amsmath,amssymb,aps,twocolumn,prb,floatfix,showpacs]{revtex4}
\usepackage[dvips]{graphicx}
\usepackage{bm}

\begin{document}

\title{$90^\circ$ Ferroelectric Domains in PbTiO$_3$: Experimental Observation and Molecular Dynamics Simulations}
\author{Kenta Aoyagi$^{1}$}
\author{Takeshi Nishimatsu$^{2}$}
\author{Takanori Kiguchi$^{2}$}
\author{Toyohiko J. Konno$^{2}$}
\author{Yoshiyuki Kawazoe$^{2}$}
\author{Hiroshi Funakubo$^{3}$}
\author{Anil Kumar$^{4,5}$}
\author{Umesh V. Waghmare$^{5}$}

\affiliation{
$^{1}$Department of Materials Science, Tohoku University, Sendai 980-8579, Japan \\
$^{2}$Institute for Materials Research (IMR), Tohoku University, Sendai 980-8577, Japan \\
$^{3}$Department of Innovative and Engineered Materials, Tokyo Institute of Technology, Yokohama 226-8503, Japan\\
$^{4}$Department of Physics and Astronomy, Rutgers University, 136 Frelinghuysen Road, Piscataway, NJ 08544-8019, \\
$^{5}$Theoretical Sciences Unit, Jawaharlal Nehru Centre for Advanced Scientific Research (JNCASR),
Jakkur, Bangalore, 560 064, India}

\begin{abstract}
We report observation of $90^\circ$ ferroelectric domain structures in transmission
electron microscopy (TEM)  of epitaxially-grown films of PbTiO$_3$. Using molecular
dynamics (MD) simulations based on first-principles effective Hamiltonian of bulk PbTiO$_3$,
we corroborate the occurance of such domains showing that it arises
as metastable states only in cooling simulations (as the temperature is lowered)
and establish characteristic stability of $90^\circ$ domain structures in PbTiO$_3$. In contrast,
such domains do not manifest in similar simulations of BaTiO$_3$.
Through a detailed analysis based on energetics and comparison between PbTiO$_3$
and BaTiO$_3$, we find that $90^\circ$ domain structures are energetically favorable only
in the former, and the origin of their stability lies in the polarization-strain coupling.
Our analysis suggests that they may form in BaTiO$_3$ due to special boundary condition
and/or defect-related inhomogeneities.
\end{abstract}

\date{\today}
\maketitle

\section{Introduction}

Ferroelectric transitions in perovskite oxides, such as BaTiO$_3$, are fluctuation
driven first-order phase transitions, and hence a state with spatially fluctuating
order parameter can readily form as a result of certain mechanical and electric
boundary conditions\cite{anilk-waghmare}. A common example of such a state is the one with
domains of ferroelectric polarization with different symmetry equivalent orientations
of order parameter that are separated by domain walls.
Indeed, many properties of perovskite ferroelectrics depend on such domain
structure, and it is being increasingly relevant
at nano-scale\cite{james-scott-nano,PhysRevB.79.144111,
Fong:S:S:E:A:F:T:Science:304:p1650-1653:2004,Paul:N:K:W:PRL:99:p077601:2007}.
Naturally, the properties of a domain wall or an interface between adjacent ferroelectric domains depend on
(a) symmetries and structural details of ferroelectric phases and (b) microscopic
couplings responsible for the ferroelectric phase transition.

Perovskite oxides such as BaTiO$_3$ and PbTiO$_3$ are representative ferroelectric materials,
although are quite different from each other in terms of their phase transitions.
While PbTiO$_3$ undergoes a single strongly first order phase transition from cubic to tetragonal structure
as temperature is lowered, BaTiO$_3$ exhibits a sequence of three relatively weaker first-order
phase transitions. A paraelectric phase of BaTiO$_3$ with cubic structure transforms into a
tetragonal ferroelectric phase at a Curie temperature, 393~K. Further cooling produces sudden
changes from a tetragonal phase to an orthorhombic phase at 278~K and from an orthorhombic phase to
a rhombohedral phase at 203~K\cite{ISI:A1951XW72000018,RevModPhys.22.221:BaTiO3}.
On the other hand, PbTiO$_3$ exhibits a phase transition from a
paraelectric cubic phase to a ferroelectric tetragonal phase
at $T_{\rm C}=763$~K and remains tetragonal down to 0 K\cite{ISI:A1950UB34500045,ISI:A1956WL85400007,ISI:A1955WB09300058}.

While the domain structures in ferroelectrics have been revealed
experimentally\cite{RevModPhys.22.221:BaTiO3,ISI:000239426600029:BaTiO3,ISI:000281059200067:BaTiO3,
ISI:000253540500036:BaTiO3,ISI:000265943200035,ISI:A19656673700004,ISI:A19646702B00043,ISI:A19626678B00035,ISI:000286219300093,Nakaki:JAP104:2008,ISI:000249474000061,ISI:000239807500005,ISI:000295824500007,ISI:000295824500017,ISI:A1995QK29400005,ISI:A1995RN01700021},
detailed  {\it in situ} experimental analysis of the domains and domain walls is quite challenging and
the temperature dependence of dynamics of domain structure is not well understood.
In the case of PbTiO$_3$,
experimental studies of domains are further more difficult as the sample needs
to be heated over its {\it high} transition temperature
$T_{\rm C}=763$~K and such heating leads to evaporation of Pb ions
changing the composition of the sample\cite{Gerson:PbTiO3:JAP:1960}.


Ferroelectric phase transitions in perovskite oxides in bulk and thin film
have been investigated by computer simulations such as
phase-field method\cite{YLLi:LQChen:APL78:2001:3878},
Monte Carlo simulations\cite{Zhong:V:R:PRB:v52:p6301:1995},
and molecular dynamics (MD) simulations\cite{Waghmare:C:B:2003}.
Recently, Nishimatsu {\it et al}.  have developed a fast and versatile MD simulator
of ferroelectrics based on first-principles effective Hamiltonian\cite{Nishimatsu:feram:PRB2008} which
can be used in systematic studies of bulk as well as thin films.
They have studied BaTiO$_3$ bulk and thin-film capacitors and obtained results
showing good agreement with experiments. Their MD simulations of BaTiO$_3$
under periodic boundary condition (PBC) for bulk did not show any domain structures,
as there is no depolarization field in the PBC of bulk. Simulations of thin-film of
BaTiO$_3$ only show $180^\circ$ domain structures\cite{Nishimatsu:feram:PRB2008},
though $90^\circ$ domain structures are widely seen in experiments\cite{ISI:000239426600029:BaTiO3,ISI:000253540500036:BaTiO3}.
One of the advantages of MD simulations compared to Monte Carlo simulations is
its ability to simulate time-dependent dynamical phenomena, e.g. MD simulation
can be used to study the evolution of ferroelectric domains as a function of time
during heating-up and cooling-down simulations.

In this paper, we report heating-up and cooling-down molecular-dynamics (MD) simulations
of bulk PbTiO$_3$ to understand our observation of $90^\circ$ domain structures in
epitaxially-grown sample of PbTiO$_3$. In Sec.~\ref{sec:expt_details}, we present
experimental details for the sample preparation and we show a
transmission electron microscope (TEM) image of $90^\circ$ domain structure in PbTiO$_3$
film. We briefly explain the first-principles effective Hamiltonian
and details of MD simulations in Sec.~\ref{sec:md_details} and we
present our results and analysis of heating-up and cooling-down MD simulations in Sec.~\ref{sec:results}.
We finally summarize our work and conclusions in Sec.~\ref{sec:summary}.

\section{Experimental Details and Observations}
\label{sec:expt_details}

\subsection{Sample preparation and Methods of TEM}

\label{sec:sample}
An epitaxial PbTiO$_3$ thick film, with film thicknesses of about 1200 nm,
was grown on the SrRuO$_3$/SrTiO$_3$ substrate at 873~K by pulsed metal organic
chemical vapor deposition (pulsed-MOCVD) method. SrRuO$_3$ was deposited on
(100) SrTiO$_3$ by rf-magnetron sputtering method. The detail of film
preparation technique is described elsewhere\cite{ISI:000085477300039,ISI:000090138700016}.
The TEM specimens were prepared with focused ion beam (FIB) micro-sampling technique.
Damage layers, introduced during FIB microfabrication, were removed by low-energy
Ar ion milling at 0.3~kV. JEM-2000EXII was used for TEM observations.
TEM observations were performed at room temperature.

\subsection{Observed TEM image}

\label{sec:image}
\begin{figure}
\centering
\includegraphics[width=88mm]{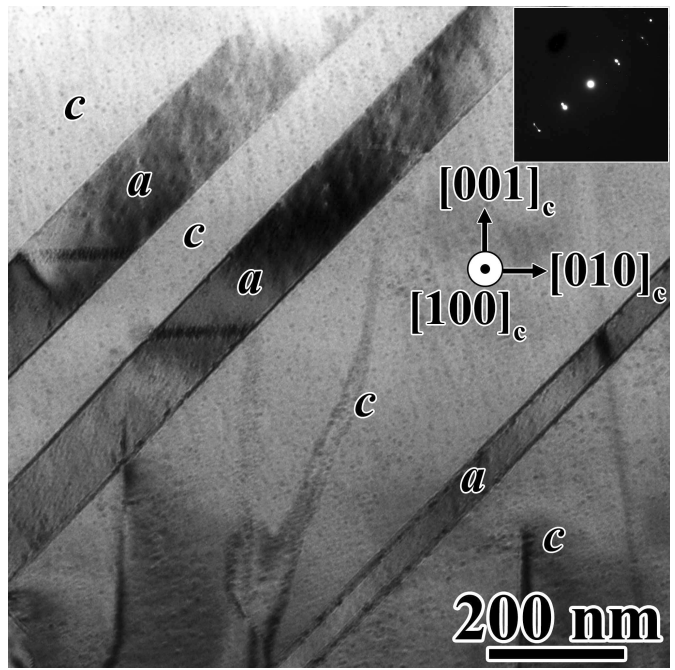}
\caption{A bright-field TEM image of a PbTiO$_3$ thick film taken with an electron
incident parallel to the [100] of PbTiO$_3$. Scripts {\it a} and {\it c} in
this figure denote {\it a}-domain and {\it c}-domain, respectively.}
\label{fig:BFTEM}
\end{figure}

\begin{figure}
\centering
\includegraphics[width=88mm]{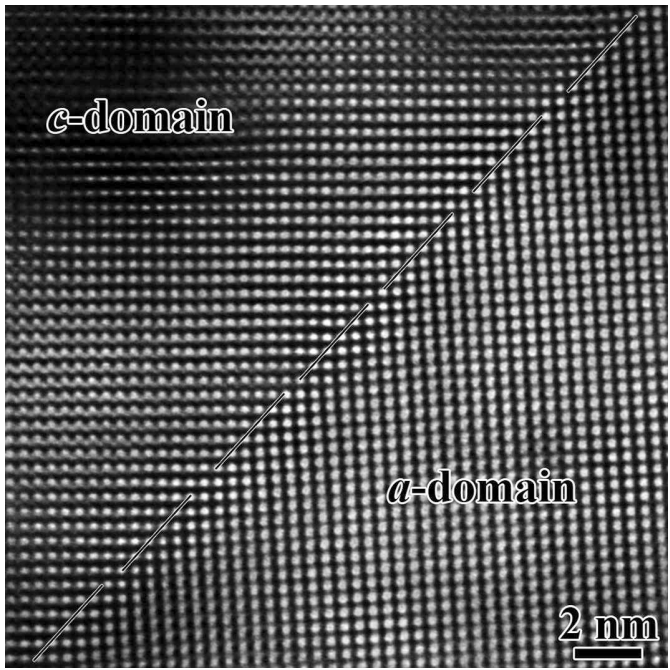}
\caption{A high-resolution TEM image of a PbTiO$_3$ thick film taken with an electron
incident parallel to the [100] of PbTiO$_3$. The $90^\circ$ domain boundary is shown by the chain line.}
\label{fig:HREM}
\end{figure}

In Fig.~\ref{fig:BFTEM}, we show a bright field TEM image of a PbTiO$_3$ thick
film, taken with the incident electron beam parallel to the [100] axis of the
PbTiO$_3$. The inset in the image is the corresponding selected-area electron
diffraction pattern. This bright-field TEM image is taken with the scattering
vector $g_{0\bar{1}\bar{1}}$ excited. Here, $a$- and $c$-domains are
those with polarization along $a$ and $c$ axes of the PbTiO$_3$ parallel and perpendicular to the
substrate, respectively. This domain configuration is typical and very
commonly seen for tetragonal PbTiO$_3$ films. The domain size is about
50-200 nm. Such $90^\circ$ domain configurations, similar to that in
Fig.~\ref{fig:BFTEM}, have been also observed in
BaTiO$_3$\cite{ISI:000239426600029:BaTiO3,ISI:000253540500036:BaTiO3}.
Boundaries between {\it a}- and {\it c}-domains are seen as black lines, as the
TEM sample is slightly tilted. From this image, we can
not discuss the width of domain walls between {\it a}- and {\it c}-domain.
High-resolution TEM observation has revealed that the width of $90^\circ$
domain walls is $1.0\pm0.3$~nm\cite{doi:10.1080/01418619508244477}.
To this end, high-resolution TEM observation was conducted in order to reveal
the width of domain walls between {\it a}- and {\it c}-domain. Fig.~\ref{fig:HREM} shows
the high-resolution TEM image of PbTiO$_3$ film, indicating that the width of
domain walls is corresponding to 1 or 2 unit cells.

Before our computational study, it should be worth mentioning that
$90^\circ$ domains have been often observed in both BaTiO$_3$ and PbTiO$_3$,
and are both ferroelectric and ferroelastic in nature.
Typically, $90^\circ$ domains are formed in epitaxial ferroelectric and ferroelastic films
in order to relax the strain resulting from lattice mismatch with the substrate
at and below $T_{\rm C}$.\cite{Pompe:APL74:p6012:y1993,ISI:A1994NW31800068}
They nucleate at misfit dislocations formed above $T_{\rm C}$.
Their growth is accompanied with the introduction of
the additional dislocation perpendicular to the misfit dislocations and the
dissociation of the dislocations into two pairs of partial dislocations
around an anti-phase boundary\cite{ISI:000295824500017}.

\section{Molecular Dynamics Simulations}
\label{sec:md_details}

\subsection{Effective Hamiltonian}
\label{sec:Eff_Hamiltonian}

Heating-up and cooling-down molecular-dynamics (MD) simulations
are performed using first-principles effective Hamiltonian
\cite{King-Smith:V:1994,Zhong:V:R:PRB:v52:p6301:1995,Waghmare:C:B:2003,Nishimatsu:feram:PRB2008,Nishimatsu.PhysRevB.82.134106},

\begin{multline}
  \label{eq:Effective:Hamiltonian}
  H^{\rm eff}
  = \frac{M^*_{\rm dipole}}{2} \sum_{\bm{R},\alpha}\dot{u}_\alpha^2(\bm{R})
  + \frac{M^*_{\rm acoustic}}{2}\sum_{\bm{R},\alpha}\dot{w}_\alpha^2(\bm{R})\\
  + V^{\rm self}(\{\bm{u}\})+V^{\rm dpl}(\{\bm{u}\})+V^{\rm short}(\{\bm{u}\})\\
  + V^{\rm elas,\,homo}(\eta_1,\ldots\!,\eta_6)+V^{\rm elas,\,inho}(\{\bm{w}\})\\
  + V^{\rm coup,\,homo}(\{\bm{u}\}, \eta_1,\ldots\!,\eta_6)+V^{\rm coup,\,inho}(\{\bm{u}\}, \{\bm{w}\})~,
\end{multline}

where $\bm{u}=\bm{u}(\bm{R})$ and $\bm{w}=\bm{w}(\bm{R})$ are, respectively,
the local dipolar  displacement vector and the local acoustic displacement vector
of the unit cell at $\bm{R}$ in a simulation supercell. $\alpha (=x,y,z)$ is the Cartesian directions.
Braces $\{\}$ denote a set of $\bm{u}$ or $\bm{w}$ in the supercell. $\eta_1,\ldots\!,\eta_6$ are
the homogeneous strain components.
$M^*_{\rm dipole}$ and $M^*_{\rm acoustic}$ are the effective masses for $\bm{u}$ and $\bm{w}$,
therefore, first two terms in Eq.~(\ref{eq:Effective:Hamiltonian}) are kinetic energies of them.

$V^{\rm self}$, $V^{\rm dpl}$, $V^{\rm short}$, $V^{\rm elas,\,homo}$,
$V^{\rm elas,\,inho}$, $V^{\rm coup,\,homo}$, and $V^{\rm coup,\,inho}$ are
a local-mode self-energy, a long-range dipole-dipole interaction,
a short-range interaction, a homogeneous elastic energy,
an inhomogeneous elastic energy,
a couping between $\{\bm{u}\}$ and $\eta_1,\ldots\!,\eta_6$, and
a couping between $\{\bm{u}\}$ and $\{\bm{w}\}$, respectively.
More detailed explanation of symbols in the effective Hamiltonian
can be found in Refs.~\onlinecite{Nishimatsu:feram:PRB2008}
and \onlinecite{Nishimatsu.PhysRevB.82.134106}.
We take all the parameters of the first-principles effective Hamiltonian for PbTiO$_3$ from the
earlier work\cite{Waghmare:R:1997PRB}. However, the form of the effective Hamiltonian
we use in our simulations\cite{Nishimatsu:feram:PRB2008,Nishimatsu.PhysRevB.82.134106} is slightly different
from the one used to get the parameters in Ref.~\onlinecite{Waghmare:R:1997PRB}.
The new parameters can be easily derived from the previous ones.
We list values of all the parameters used in our simulations and how they are related to
the previous work \cite{Waghmare:R:1997PRB} in Table~\ref{tab:parameters}.

\begin{table}
\caption{The parameters of first-principles effective Hamiltonian for PbTiO$_3$
used in our simulations are given in the second column and how these parameters
are related with the parameters in the previous work\cite{Waghmare:R:1997PRB} are
shown in the third column.}
\label{tab:parameters}
\centering
\begin{tabular}{|c|c|c|}
\hline
parameters                  & value        &  relation              \\
\hline
$a_0$     [\AA]             &  3.969        & a$_{0}$                \\
$B_{11}$  [eV]              &  117.9        & $C_{11}$               \\
$B_{12}$  [eV]              &   51.6        & $C_{12}$               \\
$B_{44}$  [eV]              &  137.0        & $C_{44}$               \\
\hline
$B_{1xx}$ [eV/\AA$^2$]      &$-114.02$      & 2($g_{0}$+$g_{0}$)     \\
$B_{1yy}$ [eV/\AA$^2$]      & $-13.67$      & 2$g_{0}$               \\
$B_{4yz}$ [eV/\AA$^2$]      & $-22.67$      &  $g_{2}$               \\
$\alpha$  [eV/\AA$^4$]       & $27.83$      & $B+C$                  \\
$\gamma$  [eV/\AA$^4$]       &$-34.48$      & $-2C$                  \\
$k_1$     [eV/\AA$^6$]       &$-42.42$      &  $D$                   \\
$k_2$     [eV/\AA$^6$]       &  0           &                        \\
$k_3$     [eV/\AA$^6$]       &  0           &                         \\
$k_4$     [eV/\AA$^8$]       & 156.43       & $E$                     \\
\hline
$m^*$     [amu]              & 100.0        &                         \\
$Z^*$     [e]                & 10.02        & $Z^*$                   \\
$\epsilon_\infty$            &  8.24        & $\epsilon_\infty$        \\
$\kappa_2$[eV/\AA$^2$]       & 1.170        & $A$                      \\
$j_1$     [eV/\AA$^2$]       &$-1.355$      & $2a_{T}$                 \\
$j_2$     [eV/\AA$^2$]       & $4.986$      & $2a_{L}$                 \\
$j_3$     [eV/\AA$^2$]       & $0.222$      & $b_{l}+b_{t1}$            \\
$j_4$     [eV/\AA$^2$]       &$-0.018$      & $2b_{t2}$                 \\
$j_5$     [eV/\AA$^2$]       &$ 0.398$      & $b_{l}-b_{t1}$            \\
$j_6$     [eV/\AA$^2$]       &$-0.083$      & $\frac{2(c_l+2c_t)}{3}$   \\
$j_7$     [eV/\AA$^2$]       &$-0.204$      & $\frac{2(c_l-2c_t)}{3}$   \\
\hline
\end{tabular}
\end{table}

\subsection{Simulation Details}
\label{sec:simulations}

Heating-up and cooling-down MD simulations
are performed with our original MD code {\tt feram} (\url{http://loto.sourceforge.net/feram/}).
Details of the code can be found in Ref.~\onlinecite{Nishimatsu:feram:PRB2008}.
Temperature is kept constant in each temperature step in the canonical ensemble
using the Nos\'e-Poincar\'e thermostat.\cite{Bond:L:L:JComputPhys:151:p114-134:1999}
This simplectic thermostat is so efficient that we can set the time step to $\Delta t=2$~fs.
In our present MD simulations, we thermalize the system for 20,000 time steps,
after which we average the properties for 20,000 time steps. We used a supercell of
system size $N = L_x\times L_y\times L_z = 32 \times 32 \times 32$
and small temperature steps in heating-up ($+1$~K/step) and cooling-down
($-1$~K/step) simulations. The heating-up simulation from 100~K to 900~K is
started from an $z$-polarized initial configuration generated randomly:
$\langle u_x \rangle = \langle u_y \rangle = 0$, $\langle u_z \rangle = 0.33~{\rm \AA}$,
$\langle u_x^2 \rangle - \langle u_x \rangle^2=\langle u_y^2 \rangle - \langle u_y \rangle^2=(0.045~{\rm \AA})^2$, and
$\langle u_z^2 \rangle - \langle u_z \rangle^2=(0.021~{\rm \AA})^2$,
where brackets denote $\bm{R}$-average in supercell $\langle u_\alpha \rangle = \frac{1}{N}\sum_{\bm{R}} u_\alpha(\bm{R})$.
The cooling-down one from 900~K to 100~K is
started from random paraelectric initial configuration:
$\langle u_x \rangle = \langle u_y \rangle = \langle u_z \rangle = 0$ and
$\langle u_\alpha^2 \rangle - \langle u_\alpha \rangle^2=(0.15~{\rm \AA})^2$.

\section{Results and Discussion}
\label{sec:results}

\begin{figure}
\centering
\includegraphics[width=88mm]{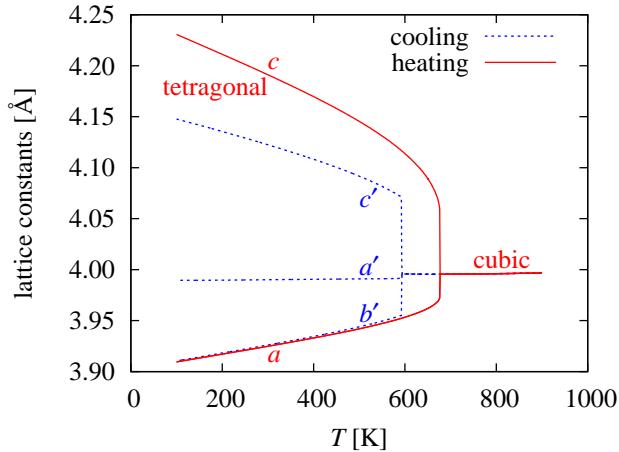}
\caption{(Color online) Simulated temperature dependence of lattice constants of PbTiO$_3$
in heating-up (red solid lines) and cooling-down (blue dashed lines) molecular-dynamics simulations.}
\label{fig:T-dep}
\end{figure}

From the temperature dependence of averaged lattice constants (shown in Fig.~\ref{fig:T-dep}),
a tetragonal-to-cubic ferroelectric-to-paraelectric
phase transition is clearly observed in the heating-up simulation at 677~K. However, a
strange behavior in lattice constants is found
in the cooling-down simulation at T=592 K. The average temperature of these two transition temperatures
(634 K) is in good agreement with the earlier Monte Carlo simulations and slightly lower than the
 experimental value $T_{\rm C}=763$~K. Indeed, the observation of an orthorhombic phase
 during cooling-down simulations is intriguing.

\begin{figure*}
\centering
\includegraphics[width=140mm]{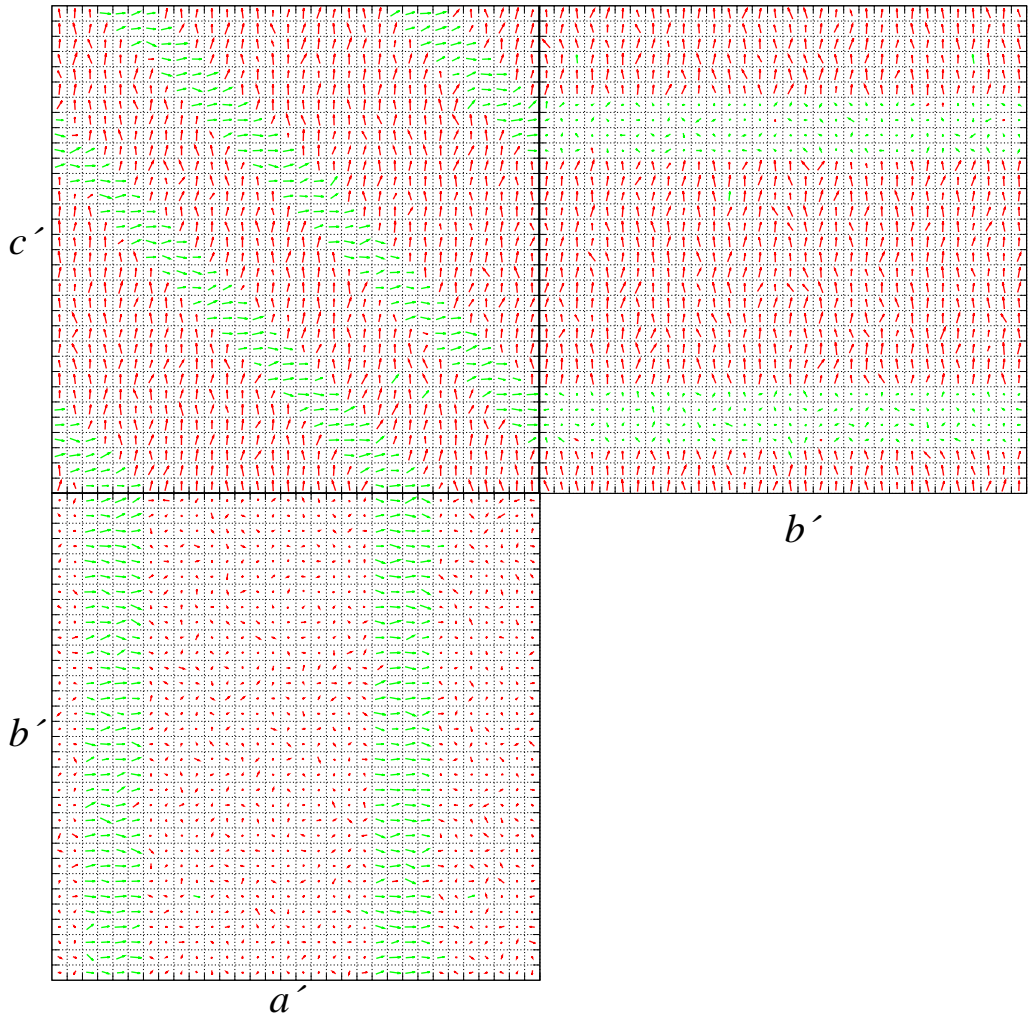}
\caption{(Color online) Snapshot of three ``sides'' of the $32 \times 32 \times 32$ supercell at $T=300$K in a
cooling-down simulation of PbTiO$_3$. Dipole moments of each site are projected
onto the side planes and indicated with arrows. Dipoles of $u_z>0.2$\AA\ are indicated
with red color. Dipoles of $u_z\leq 0.2$\AA\ are indicated with green.
Crystalline directions are indicated with $a'$, $b'$, and $c'$ as indicated in Fig.~\ref{fig:T-dep}.}
\label{fig:domain}
\end{figure*}

To understand this interesting behaviour of lattice constants in cooling-down simulations, we
perform a detailed analysis of the configurations (snapshots) during our MD simulations.
From a snapshot of dipoles in the supercell (shown in Fig.~\ref{fig:domain}),
we find that the apparently orthorhombic nature of the phase
is due to a $90^\circ$ domain structure.
Although the 4~unit~cell $= 1.6$~nm
of the domain size is much smaller than experimentally observed ones as shown in
Sec.~\ref{sec:image},
the width of a simulated domain wall estimated to be $\sim~1$~unit cell
is in good agreement with our experiment.
Each domain has
the $a=b<c$ tetragonal structure of PbTiO$_3$, but their average value in whole crystal gives
smaller $c'$ than $c$ and larger $a'$ than $a$. The lattice constant $b'$ has
almost the same values as $a$, because the polar directions of two kind of domains are
perpendicular to the $b'$-axis. It should be noted that this domain structure is found in
MD simulations in bulk under periodic boundary condition (PBC), but we have not
simulated thin films.
Under the PBC, there is no depolarization field inside the bulk.
Moreover, this domain structure can be easily reproduced in cooling-down simulations
from any random paraelectric initial configurations and any seeds
for the pseudo random number generator\cite{Marsaglia:Tsang:RNG:2004}. There was no evidence
for such unusual behavior in simulations of bulk BaTiO$_3$ \cite{Nishimatsu:feram:PRB2008,Nishimatsu.PhysRevB.82.134106}.

To understand the reason of stability of the $90^\circ$ domain structure seen here,
even in bulk PbTiO$_3$,
we compare ``total energy surfaces'' between single and $90^\circ$ domain structures. The total
energy surface of single domain structure with $[001]$ polarization is the same as in
Refs.~\onlinecite{King-Smith:V:1994} and \onlinecite{Nishimatsu:feram:PRB2008}. For the total
energy surface of $90^\circ$ domain structure, we focus on a snapshot of the supercell at 300K
shown in Fig.~\ref{fig:domain} and represent it with $\{\bm{u}_{90^\circ\!,300{\rm K}}(\bm{R})\}$.
We now obtain a sequence of configurations by multiplying a
factor $\frac{u}{\langle u \rangle_{90^\circ\!,300{\rm K}}}$ for all $\bm{R}$

\begin{equation}
\label{eq:u90}
\bm{u}(\bm{R})=\frac{u}{\langle u \rangle_{90^\circ\!,300{\rm K}}}\bm{u}_{90^\circ\!,300{\rm K}}(\bm{R})~,
\end{equation}

 and compute total energy as a function of $u$, where $\langle u \rangle_{90^\circ\!,300{\rm K}}$
is the averaged length of dipoles in the 300~K snapshot

\begin{equation}
\label{eq:u-redefined}
\langle u \rangle_{90^\circ\!,300{\rm K}} = \frac{1}{N}\sum_{\bm{R}}|\bm{u}_{90^\circ\!,300{\rm K}}(\bm{R})|=0.32~{\rm \AA}~.
\end{equation}

\begin{figure}
\centering
\includegraphics[width=80mm]{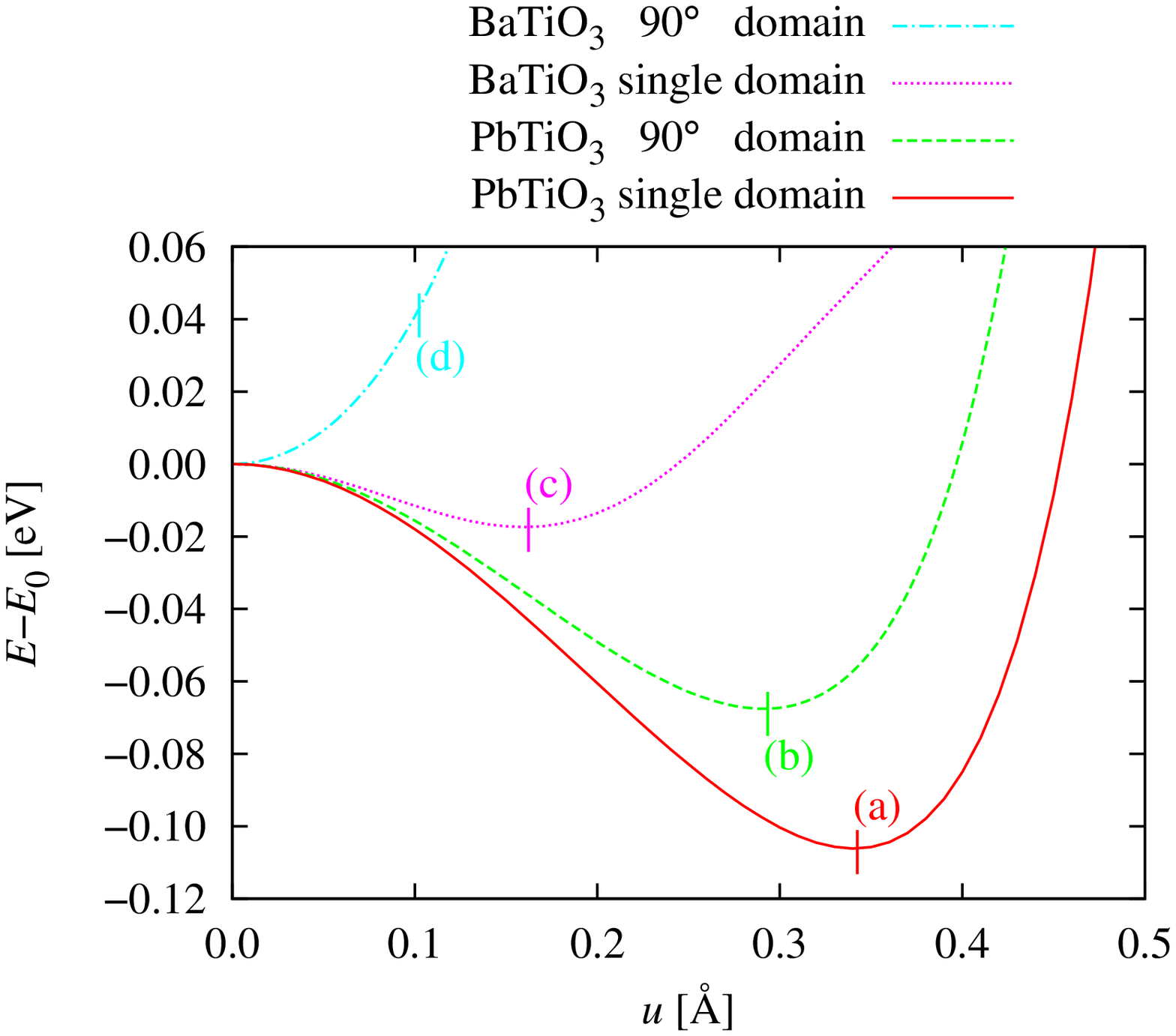}
\caption{(Color online) Calculated total energy surfaces for
  PbTiO$_3$ single domain (solid line),
  PbTiO$_3$ $90^\circ$ domain (dashed line),
  BaTiO$_3$ single domain (dotted line), and
  BaTiO$_3$ $90^\circ$ domain (chain line).
}
\label{fig:surfaces}
\end{figure}

Calculated total energy surfaces of single and $90^\circ$ domain structures for PbTiO$_3$
are shown in Fig.~\ref{fig:surfaces} with solid and dashed lines, respectively.
For comparison, those for BaTiO$_3$ are also plotted assuming the same $90^\circ$ domain structure
by using the set of parameters in $H^{\rm eff}$ listed in Ref.~\onlinecite{Nishimatsu.PhysRevB.82.134106}.
While the $90^\circ$ domain structure of PbTiO$_3$ exhibits a minimum at $u\neq 0$,
that of BaTiO$_3$ costs energy. This is why the $90^\circ$ domain structures can be found in simple
cooling-down simulations of PbTiO$_3$, but not in those of BaTiO$_3$.


Minima are indicated with (a)--(c) in Fig.~\ref{fig:surfaces}. To uncover the origin of this
contrasting behaviour, interaction energy terms at the minimums are listed in TABLE~\ref{tab:V}.
For the BaTiO$_3$, because there is no minimum, interaction energies of configuration at (d)
in Fig.~\ref{fig:surfaces} (of $u=0.10$~\AA ) are listed.

\begin{table*}
\caption{Comparison of interaction energies $V^i$ in the
  effective Hamiltonian of Eq.~(\ref{eq:Effective:Hamiltonian}) for
  two kinds of domain states of PbTiO$_3$ and BaTiO$_3$.
  $V^{\rm harmonic}$ is the sum of $V^{\rm dpl}$, $V^{\rm short}$, and the harmonic terms in $V^{\rm self}$.
  $V^{\rm unharmonic}$ is the unharmonic terms in $V^{\rm self}$.
  Unit of energy is eV.}
\label{tab:V}
\centering
\begin{tabular}{|l|rrr|rrr|}
\hline
$x$ of     & PbTiO$_3$ & PbTiO$_3$  && BaTiO$_3$  & BaTiO$_3$ &\\
interaction&(a) single &(b) $90^\circ$
                                    &&(c) single  &(d) $90^\circ$&\\
energy $V^x$& domain   & domain     &$\Delta V^i$& domain     & domain    &$\Delta V^i$\\
\hline
harmonic   &$-0.22106$&$-0.14393$&$+0.07713$&$-0.03886$&$ 0.03541$&$+0.07427$\\
unharmonic &$ 0.33435$&$ 0.17249$&$-0.16186$&$ 0.04754$&$ 0.00695$&$-0.04059$\\
elas,homo  &$ 0.21946$&$ 0.04441$&$-0.17505$&$ 0.02609$&$ 0.00131$&$-0.02478$\\
elas,inho  &$ 0.00000$&$ 0.05171$&$+0.05171$&$ 0.00000$&$-0.00262$&$-0.00262$\\
coup,homo  &$-0.43891$&$-0.08881$&$+0.35010$&$-0.05218$&$ 0.00089$&$+0.05307$\\
coup,inho  &$ 0.00000$&$-0.10342$&$-0.10342$&$ 0.00000$&$-0.00178$&$-0.00178$\\
total      &$-0.10616$&$-0.06755$&$+0.03861$&$-0.01741$&$ 0.04016$&$+0.05757$\\
\hline
\hline
$u$ [\AA]  &0.34      &0.29      &&0.16      & 0.10    &\\
\hline
\end{tabular}
\end{table*}

From TABLE~\ref{tab:V}, It is clear that the energy losses in $V^{\rm harmonic}$,
$V^{\rm elas,\,inho}$, and $V^{\rm coup,\,homo}$ in forming the $90^\circ$ domain structure
in PbTiO$_3$ are compensated by the energy gains in $V^{\rm unharmonic}$, $V^{\rm elas,\,homo}$, and $V^{\rm coup,\,inho}$.
In contrast, such recovery is not sufficient in BaTiO$_3$ to form a $u\neq 0$ minimum or non trivial 90$^{\circ}$ domain
structures. In stabilization of the 90$^{\circ}$ domain structures in PbTiO$_3$, our analysis conclusively highlights
the role of two microscopic interactions: (a) lower elastic energy cost arising from the smaller strain from compensation along
$c$ and $a$ axis, and (b) inhomogeneous (local) strain coupling with polarization at the domain wall. The latter does not
contribute much to transition behavior in the bulk, but has rather significant impact on the properties of domain wall.
Noting that a $90^\circ$ domain structure is not energetically favorable in BaTiO$_3$
simulated as a perfect bulk crystal and coupling of soft modes with higher energy modes is weak,
we believe that the experimentally observed $90^\circ$ domain structures
in BaTiO$_3$ is most likely due to inhomogeneities in the samples and/or the specific electric and
mechanical boundary conditions.

\section{Summary}
\label{sec:summary}

In this article, TEM observation of PbTiO$_3$ thick films revealed $90^\circ$ domain structures, 
which have been often observed in BaTiO$_3$. The domain size perpendicular to domain boundaries was 50--200~nm. 
The width of domain wall was corresponding to 1 or 2 unit cell.

We also have performed heating-up and cooling-down MD simulations of PbTiO$_3$. In cooling-down simulation,
$90^\circ$ domain structure is found to form spontaneously.
By comparing ``total energy surfaces'' of single and $90^\circ$ domain structures,
we understand that a $90^\circ$ domain structure
is metastable in bulk PbTiO$_3$, but not in bulk BaTiO$_3$.
The origin of this contrast is traced to significantly larger
polarization-stran coupling in PbTiO$_3$.
Hence,
while $90^\circ$ domain structures can form spontaneously in PbTiO$_3$,
they seem to arise in BaTiO$_3$
mostly from special boundary conditions
and/or defect-related inhomogeneities.

\section*{Acknowledgments}
This work was supported by
Japan Society for the Promotion of Science (JSPS) through KAKENHI 23740230 and 21760524.
Computational resources were provided by the Center for Computational Materials Science,
Institute for Materials Research (CCMS-IMR), Tohoku University. We thank the staff at
CCMS-IMR for their constant effort. This study was also supported by the Next Generation
Super Computing Project, Nanoscience Program, MEXT, Japan. UVW acknowledges an IBM faculty
award grant in supporting some of his work.

\bibliography{BaTiO3,PbTiO3,Pulsed-MOCVD,ferroelectrics,MD,marsaglia_tsang_uni64}
\end{document}